\documentstyle[multicol,aps,prl,epsfig]{revtex}

\begin{document}

\draft

\wideabs{
\title{Search for Magnetic Field Induced Gap in a High-$T_c$ Superconductor}
\author{G. Mih\'aly$^{\#}$, A. Halbritter$^{\#}$, L. Mih\'aly$^{\ast}$, and L. Forr\'o}
\address{Department of Physics, EPFL, CH-1015 Lausanne }
\date{\today}
\maketitle

\begin{abstract}
Break junctions made of the optimally doped high temperature superconductor
Bi$_{2}$Sr$_{2}$Ca$_{2}$CuO$_{8}$ with $T_{c}$ of 90~K has been investigated
in magnetic fields up to 12~T, at temperatures from 4.2~K to $T_{c}$. The
junction resistance varied between 1~k$\Omega $ and 300~k$\Omega$. The
differential conductance at low biases did not exhibit a significant
magnetic field dependence, indicating that a magnetic-field-induced gap
(Krishana {\it et al.}, Science {\bf 277} 83 (1997)), if exists, must be smaller
than 0.25~meV.
\end{abstract}

}

Much of what we know about the electronic states in high-$T_{c}$
superconductors has been learned by using tunneling devices different from
the metal-insulator-metal layer junctions so successful in exploring
traditional superconductivity \cite{wolf}.
Superconductor-insulator-superconductor (SIS) tunneling on break junctions
provided one of the first clear indications for the failure of a fully
gapped $s$-wave density of states (DOS) in these materials \cite
{mandrus,mandrus2,hartge}. Optimally doped samples were studied by tunneling
in great detail \cite{hancotte,romano,tao}. More recently, the oxygen doping
dependence of the gap has been investigated on SIS junctions created by
proper manipulation of a normal metal point contact \cite{miyakawa}.
Superconductor-insulator-normal metal (SIN) junctions were used very
successfully in scanning tunneling spectroscopy 
studies \cite{dewilde,renner}, and in point contact 
measurements \cite{huang,kane}. Although many of
these junctions are much less controlled than the traditional metal oxide
layer junctions, there is a reasonable level of consistency between the
various techniques, indicating that the features observed are intrinsic to
the materials.

The present study was motivated by a recent report of Krishana 
{\it et al.}\ on the magnetic field dependence of the 
thermal conductivity \cite{krishana}, and
by theoretical arguments about the behavior of $d$-wave superconductors in
magnetic field \cite{balatsky,laughlin,janko}. The apparent non-analytical
behavior reported by Krishana {\it et al.}\ raised the 
intriguing possibility of a
magnetic field induced instability of the $d$-wave superconducting state with
the appearance of a complete gap at temperatures below 20~K and magnetic
fields of the order of 1~T. Theoretical foundations in terms of mixing a
(complex) $d_{xy}$ component to the $d_{x^{2}-y^{2}}$ state were suggested
by Balatsky \cite{balatsky} and Laughlin \cite{laughlin}.

We performed break junction tunneling measurements in magnetic fields, with
the goal of testing these conjectures directly. The tunneling device used in
this work is an advanced version of an earlier one \cite{mandrus}, used most
recently in the study of the superconducting gap of Rb$_{3}$C$_{60}$ \cite
{koller}. In short, a very thin optimally doped 
Bi$_{2}$Sr$_{2}$Ca$_{2}$CuO$_{8}$ (BSCCO) single crystal 
of $T_{c}=90$~K was mounted on a flexible
support, and contacted with gold wires. The sample was cooled in He
atmosphere, and a break junction was created {\it in situ} by bending the
support. (A similar method has been employed for Al and Nb point contacts by
Scheer {\it et al.}\ \cite{scheer} and 
Muller {\it et al.}\ \cite{muller}, respectively.)
A piezoelectric rod was used for fine tuning of the junction. The tunneling
current was parallel, and the magnetic field was perpendicular to the copper
oxide planes. Early reports of point contact spectroscopy \cite{vedeneev91}
and break-junction tunneling on BSCCO superconductor in magnetic field by
Vedeneev {\it et al.}\ \cite{vedeneev94} did 
not address the same issue, they did not have sufficient resolution to 
investigate the effect of the magnetic field on the shape of the low
bias region of the conductance curves and furthermore their
data were interpreted within a thermally broadened $s$-wave symmetry BCS gap.

In zero field and at low temperatures earlier results of ours \cite
{mandrus,mandrus2,hartge} and others \cite{huang,kane} have been reproduced.
At low temperatures the zero bias conductance of the junctions was close to
zero. At finite bias voltages the differential conductance followed an
approximately quadratic bias dependence at low voltages and exhibited peaks
around $\pm$60~mV. The corresponding peak-position in the density of states
of the $d$-wave superconductor (the $d$-wave gap) 
is at $\Delta =40$~meV \cite{dwave}. This is in general 
agreement with the values reported for an
optimally doped sample \cite{mandrus,hartge,hancotte}.

If the magnetic field induces a gap $\Delta ^{\prime}$ in the density of
states, then the differential conductance is expected to change: in the low
temperature limit a fully gapped DOS results in vanishing tunneling
conductivity for voltages up to $2\Delta ^{\prime }/e$. Numerous junctions
were measured in search of this effect. In Figs.~1 and 2 two examples are
shown, representing high and low resistance junctions, respectively. In
the upper pannel of Fig.~1 the $I$-$V$ characteristics at $B=0$ and 12~T
fields are shown in the $\pm200$~mV range. The middle pannel displays
the 
corresponding condunctace curves normalized to the value at 200~mV in zero
magnetic field. The low bias part of the conductance curve is blown up in
the lower panel of Fig.~1. The curves at $B=1$, 2 and 12~T are shifted in
respect to the zero field value for the sake of a better view. 
The same parabola
(corresponding to a linear density of state) is drawn as an eyeguide over
the experimental points for scans at differenet magnetic fields. It is
evident from these curves that no change was found in the shape of the
tunneling characteristic in the temperature and magnetic field range where
the thermal conductivity anomaly was observed by Krishana 
{\it et al.}\ \cite{krishana}. The absence of magnetic field dependence of
the tunneling 
conductance at 4.2~K is illustrated in Fig.~2 for the case of a low
resistance junction. Similar conclusions were reached for temperatures up to
30~K on all the samples investigated.

\begin{figure}
\noindent
\centerline{\includegraphics[angle=-90,width=0.82\linewidth]{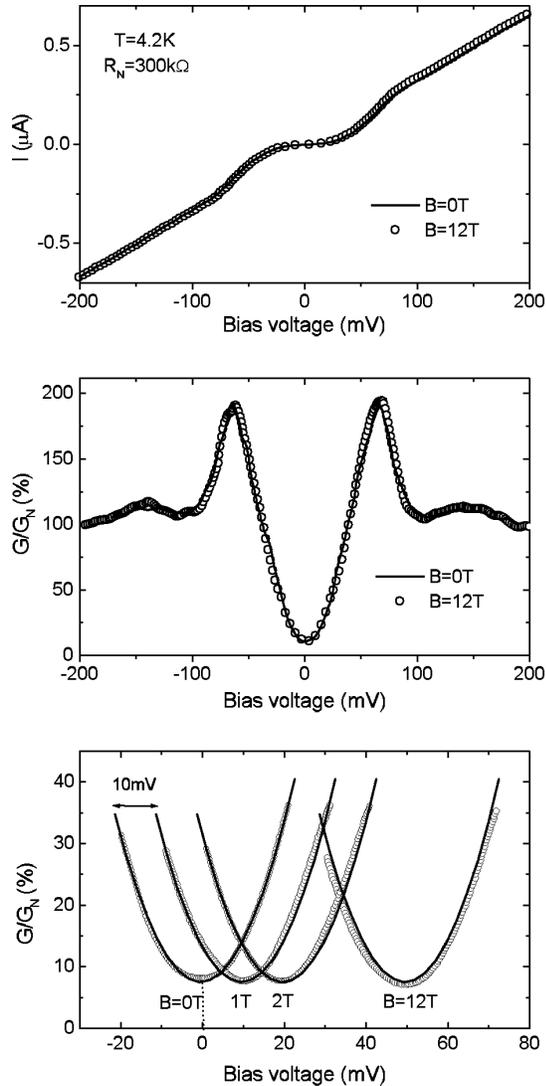}}
\vspace{1truemm}
\caption{\it Upper panel: Tunneling current vs. voltage ($I$-$V$) characteristics
for SIS break junction at 4.2~K on an optimally doped BSCCO crystal in zero
and 12~T magnetic field. Middle panel: The corresponding differential
conductance curves obtained by numerical derivation of the $I$-$V$
characteristics of the upper panel. The conductance is normalized to its
zero field value at the bias voltage of 200~mV. The resistance of the
junction is $R=300$~k$\Omega $ at 200~mV. Lower panel: the low-bias portion
of the differential 
conductance curves, measured in various magnetic fields (circles). The
curves in non-zero magnetic fields are shifted for a better view. The solid
line, supplied as a reference, represents the {\it same} quadratic function.
A complete gap of magnitude $\Delta ^{\prime }$ in the density of states
should result in a $4\Delta ^{\prime }/e$ wide region of zero conductivity
around zero bias. }
\label{fig1}
\end{figure}

It should be mentioned that the finite size of the junctions averages the
physical properties on the involved area. For a $d$-wave superconductor, a
full gap should be expected only for pure tunneling along $a$-$a$ or $b$-$b$
directions (on a microscopic scale). In these junctions the tunneling
averages the density of states for many $k$-values, and we always see a
quadratic voltage dependence of the conductance at low biases. Nevertheless,
if the magnetic field would suppress the nodes in the gap, the conductance
would be zero below the lowest gap value on the Fermi surface, no matter
how is the average done in different $k$-directions.

\begin{figure}[t]
\noindent
\includegraphics[angle=-90,width=1.0\linewidth]{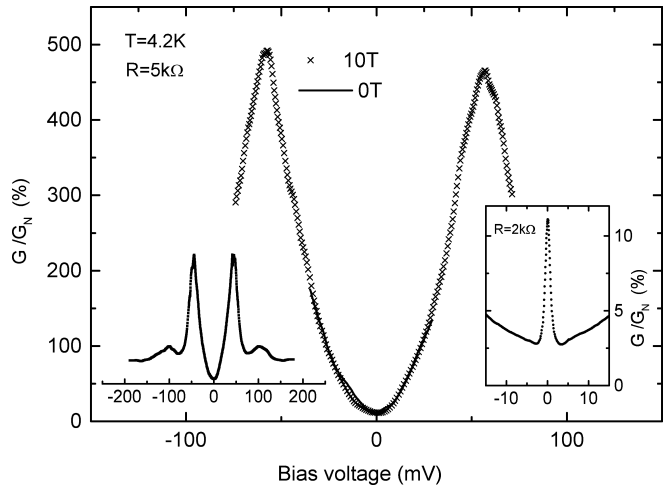}
\vspace{1truemm}
\caption{\it Main panel: Differential conductance calculated by numerical
derivation of the $I$-$V$ curve of a break junction at $T=4.2$~K. The
large-bias 
resistance of the junction is $R=5$~k$\Omega $. The continuos line denotes
the data taken in zero magnetic field. Crosses represent the data points in
high magnetic field. There is no measurable change in the tunneling
characteristic due to the application of the magnetic field. The
differential conductance at zero magnetic field over a wider range of bias
voltages is also plotted in the main panel. The inset demonstrates the
voltage resolution of the experiments: it shows the Josephson-peak observed
in a $R=2$~k$\Omega $ junction.}
\label{fig2}
\end{figure}

Josephson current is observed in low resistance junctions. In the
differential conductance, calculated from the measured currents and
voltages, this feature shows up as a sharp peak, centered at zero voltage.
Ideally, the peak should be very narrow. The width of the peak is a good
measure of our experimental resolution - an example is shown in the inset of
Fig.~2. From the half width of the curve we deduce a
voltage resolution of about $\approx 1$~mV from these measurements. Combining
the results illustrated in Figs.~1 and 2 with the resolution deduced from
the Josephson current measurements, a magnetic field induced gap of $\Delta
^{\prime }>0.25$~meV is excluded by the 
present study. (Note that a gap of $\Delta^{\prime}$ produces a 
conductivity change over the voltage range of $4\Delta^{\prime}/e$).

From a purely experimental perspective, the thermal conductivity ($\kappa $)
data of the Princeton group \cite{krishana} places a lower limit on $\Delta
^{\prime }$. According to the interpretation favored by the authors, the
absence of all field dependence in $\kappa $ is due to an exponentially
vanishing quasiparticle population - in other words it is due to a gap that
is significantly larger than the temperature. 
For example, at $B=2$~T and $T=10$~K the gap should be 
$\Delta ^{\prime }\gg 1$~meV, and it is expected to {\it increase} 
at higher magnetic fields and decreasing temperatures. This is not
compatible with our observations, in particular with the 4.2~K data shown on
the Figures.

The review of current theories reveals several of possibilities for
introducing a new energy scale to the problem. We will use a representative
magnetic field of $B=10$~T for quantitative comparison. The cyclotron
energy, in the order of $\hbar \omega _{c}\propto av_{F}{\frac{eB}{c}}$ is
about 1~meV at this field (using reasonable values of the lattice spacing $a$
and Fermi velocity $v_{F}$). A fine structure on this scale is close the
limit of the voltage resolution in the present experiment, and it can not be
entirely excluded. Janko \cite{janko} describes the states in the Abrikosov
vortices, obtaining characteristic energies in the order of magnitude of the
geometric mean of the superconducting gap and the cyclotron frequency, 
$\Delta ^{\prime }\propto \sqrt{\Delta \hbar \omega _{c}}$. The energy is in
the order of 15~meV, clearly excluded by our measurement. Finally, Laughlin 
\cite{laughlin} describes a mechanism where the new gap is $\Delta ^{\prime
}=\hbar v\sqrt{2{\frac{eB}{\hbar c}}}=5$~meV (here $\hbar v=4.5\time 10^{6}$
cm\,sec$^{-1}$ was used for the root mean square 
velocity of the $d$-wave node). A
fully gapped DOS of $\Delta ^{\prime }=5$~meV should result in reduced
conductivity over a 20~mV wide voltage range, that is clearly 
contradicted by the
experimental results shown in Figs.~1 (lower panel) and 2.

In conclusion, our tunneling measurements on break junctions place an upper
limit of $\Delta ^{\prime }\approx 0.25$~meV on the magnetic field induced
gap in BSCCO. The field induced anomaly in the thermal conductivity must
have some other explanation, possibly along the lines suggested by Aubin
{\it et al.}\ \cite{aubin}.

We thank E.\ Tuti\'{s} and B.\ Janko for valuable discussions. LM is indebted
to L. Zuppiroli for his hospitality. This project was supported by the Swiss
National Science Foundation and by OTKA T026327.

${\#}$Permanent address: Technical University of Budapest, Hungary

${\ast}$Permanent address: Department of Physics, SUNY @ Stony Brook, NY
11794, USA


\begin{references}
\bibitem{dwave}  The straightforward argument, resulting in the maximum
differential conductance of the SIS junction at $V_{peak}=2\Delta $, works
only for the isotropic $s$-wave condensate. Otherwise $V_{peak}$ can be
determined only after calculating the convolution of the density of states,
as discussed in \cite{mandrus}.
\end{references}
\end{document}